# A note on the Woods-Saxon potential

**B. Gönül and K. Köksal**

Department of Engineering Physics, University of Gaziantep, 27310, Gaziantep-Türkiye

**Abstract**

The $s-$wave Schrödinger and, to clarify one interesting point encountered in the calculations, Klein-Gordon equations are solved exactly for a single neutron moving in a central Woods-Saxon plus an additional potential that provides a flexibility to construct the surface structure of the related nucleus. The physics behind the solutions and the reliability of the results obtained are discussed carefully with the consideration of other related works in the literature. In addition, the exhaustive analysis of the results reveals the fact that the usual Woods–Saxon potential cannot be solved analytically within the framework of non-relativistic physics, unlike its exactly solvable relativistic consideration.

## 1. Introduction

There is an intimate relationship between the Jacobi polynomials and the hypergeometric function [1]. Any wavefunction expressed in terms of Jacobi polynomials can also be expressed in terms of hypergeometric functions as well. Nevertheless, in some cases it is more convenient to use these polynomials, because a wide class of solvable potentials can be found more easily if we take them as a starting point [2].

Along this line, Levai [2] investigated a simple method of constructing potentials for which the Schrödinger equation can be solved exactly in terms of special functions of mathematical physics and showed the relationship between the formalism introduced and the supersymmetric quantum mechanics [3]. The author discussed in his comprehensive article that the relation mentioned above is automatically satisfied for the ground state $(n=0)$ in the case of orthogonal polynomials. If one takes the hypergeometric function $_2F_1(\gamma,\eta,\delta;s(r))$, this condition is equivalent with $\gamma\eta = 0$ for the ground state. This is the case indeed for most of the exactly solvable potentials with solutions containing the hypergeometric function.

But there are some exceptions, for instance in the case of Woods-Saxon potential (with

$\ell = 0$), $\gamma\eta \neq 0$. There are some indications that this must be the case. For example, there is no explicit expression for $E_n$ for the Woods-Saxon potential, only a transcendental equation which has to be solved in order to obtain eigenvalues [4], although the corresponding wavefunction is known explicitly in terms of the hypergeometric fuction. Within this context, recently two works [5,6] involving different treatment techniques have been appeared in the literature to find an analytical solution for the potential of interest within the framework of non-relativistic physics. With the choice of the most frequently used type of the Jacobi polynomials with $\gamma\eta = 0$, they have arrived at consistent expressions for the energy spectrum of an extended form of the Woods-Saxon potential. Nevertheless, due to this wrong starting point leading to inevitably an improper hypergeometric function for the solutions, the analytical wavefunctions obtained in these works do not satisfy the required boundary condition in the asymptotic region. To remove this unphysical situation, Fakhri and his co-worker [5] invoked a plausible weight function into the calculations. But this suggestion increases the calculational workload and leads to cumbersome procedures for the definition of the wavefunctions.

Therefore, at this stage it is desirable to have a clean route for the introduction of reliable, practical and compact expressions leading to the exact calculation of both the energy values and wavefunctions of the Woods-Saxon potential, which is the aim of the present article.

The arrangement of this article is as follows. In the next section, a brief survey of the model used is given within the non-relativistic framework and the related formulae necessary for subsequent sections are collected. Section 3 involves the application of the formalism to the Woods-Saxon potential. For a further analysis, the same problem is also analysed within the frame of the Klein-Gordon equation, which reveals the inter-connection between the relativistic and non-relativistic calculation results obtained. To our knowledge, such a discussion has not been presented in the related literature. Finally, the results are summarized in the concluding section.

## 2. Formalism

Starting with the consideration of the Schrödinger equation $(\hbar = 2m = 1)$,

$$\frac{\Psi''(r)}{\Psi(r)} = V(r) - E \quad , \tag{1}$$

and remembering that its solutions generally take the form

$$\Psi(r) = f(r)F[s(r)],\tag{2}$$

one easily shows that the substitution of (2) into (1) yields the second-order differential equation

$$\left(\frac{f''}{f} + \frac{F''s'^2}{F} + \frac{s''F'}{F} + 2\frac{F's'f'}{Ff}\right) = V - E ,\tag{3}$$

that is reduced to the form of

$$F'' + \left(\frac{s''}{s'^2} + 2\frac{f'}{s'f}\right)F' + \left(\frac{f''}{s'^2 f} + \frac{E-V}{s'^2}\right)F = 0 .\tag{4}$$

The above equation is in the form of the most familiar second-order differential equations to the hypergeometric type [1] with an appropriate coordinate transformation $s = s(r)$

$$F''(s) + \frac{\tau(s)}{\sigma(s)}F'(s) + \frac{\lambda}{\sigma(s)}F(s) = 0 ,\tag{5}$$

where $\tau$ and $\sigma$ at most first and second degree polynomials, respectively, while $\lambda$ is a constant related to the energy values. It is well known that many of the special functions represent solutions to differential equations of the form in (5) where the functions $\tau/\sigma$ and $\lambda/\sigma$ are well defined for any particular function [1,2].

Thus, the comparison of Eqs. (4) and (5) gives

$$\frac{s''}{s'^2} + 2\frac{f'}{s'f} = \frac{\tau}{\sigma} \quad , \quad \frac{f''}{s'^2 f} + \frac{E-V}{s'^2} = \frac{\lambda}{\sigma} .\tag{6}$$

From our earlier work [7], the energy and potential terms in (6) can be decomposed in two pieces, which provides a clear understanding for the individual contributions of the $F$ and $f$ terms to the whole of the solutions, such that $E - V = (E_F + E_f) - (V_F + V_f)$. Therefore, the second equality in (6) is transformed to a couple of equation

$$\frac{f''}{f} = V_f - E_f \quad , \quad -\frac{\lambda}{\sigma}s'^2 = V_F - E_F ,\tag{7}$$

where $f$ can be expressed in an explicit form due to the first part in (6)

$$f(r) \approx (s')^{-1/2} \exp\left[\frac{1}{2}\int^{s(r)}(\tau/\sigma)ds\right] .\tag{8}$$

We note that the present simple formalism does not include a restriction, such as $\tau' \prec 0$ as in the Nikiforov-Uvarov method [8] which was used in [6]. This is one of the significant point in this article which employs another type of the Jacobi polynomials with $\tau' \prec 0$, unlike the works in [5,6], to initiate the present calculations.

## 3. Application

The shell model has been quite successful in describing nuclear structure, and the single-particle potential is important as the basic element of the shell model. The single-particle potential is often assumed to be in a simple functional form, e.g. the Woods-Saxon form [9], and the parameters in it are assumed to vary smoothly with $Z$ and $N$ [10]. If we take the single-particle potential as the average potential, Hartree-Fock calculations [11] might be more appropriate. In this kind of calculation, however, there are two problems: One is that the calculations take much computer time, and the other more serios problem is what internucleonic force we should use. In connection with this, for example, one often adopts the harmonic oscillator potential in shell model calculations for both spherical and deformed nuclei. However, due to the incorrect asymptotic property of the harmonic oscillator wave functions, the expansion in the localized such basis is not appropriate for the description of drip line nuclei. Therefore, in practice, we cannot expect the Hartree-Fock calculations to be definitely superior to the assumption of a simple form, to construct a spherical single-particle potential which is more accurate and applicable to a wider nuclidic region than attained before.

The central component of such interactions is in general considered as an extension of the Woods-Saxon potential which is the main interest in this article. We hope that the exact treatment of this simplified interaction potential presented in this section, for the zero angular momentum within the relativistic and non-relativistic frames, would initiate some attempts to put forward explicit expressions for the analytical solution of more complex form of the Woods-Saxon potential involving angular momentum barrier and the non-central parts of the interaction for not only the spherical nuclei but the deformed ones as well. In addition, the present discussion also would be helpful for the researchers dealing with the similar topics since the various modifications of the potential of interest have been considered for describing metallic clusters [12] in a successful way and widely used in molecular physics to search laser induced reactions, photodissociation and chemical reactions [13].

## 3.1. Non-relativistic consideration

From the differential equation (Eq. 5) of the Jacobi polynomials $F = (1-s)^\alpha (1+s)^\beta P_n^{(\alpha,\beta)}$ [1],

$$\sigma = 1 - s^2 \;, \quad \tau = (\alpha - \beta) + (\alpha + \beta - 2)s \;, \quad \lambda = (n+1)(n+\alpha+\beta), \quad n = 0,1,2,\ldots \;, \tag{9}$$

which can be safely used in Eq. (7) without causing any problem unlike the Nikiforov-Uvarov method [6,8] for which this polynomial is not suitable for the present calculations since $\tau' \succ 0$. To start, we also need to recall that for the real parameters $\alpha, \beta \succ -1$ which serves a testing ground for the reliability of the present calculations.

As we have a constant $(E_F, E_f)$ on the right-hand sides of two equalities in Eq. (7), there must be at least one term on the left-hand sides of these equations, from which a constant arises. In the most general case this must be one of the terms containing the parameters $n$, $\alpha$ and $\beta$ of the Jacobi polynomials [2]. Therefore, we start with a choice of

$$\frac{s'}{1-s^2} = a, \tag{10}$$

where $a$ is a positive constant and the corresponding solution is $s = (e^{2ar} - 1)/(e^{2ar} + 1)$. In order to have more realistic results based on physics, one needs to set $r \equiv r - R$ with $R \,(= r_0 A^{1/3})$ being the radius of the nucleus where the average interaction takes place. Hence,

$$s(r) = \frac{e^{2a(r-R)} - 1}{e^{2a(r-R)} + 1}, \tag{11}$$

which smoothly satisfies Eq. (10). As a result of this consideration, $2a$ term gains a physical meaning such that $2a = 1/a_0$ with $a_0$ being the diffuseness parameter that determines the thickness of a surface layer in which the potential falls off from outside $(V = 0)$ to its maximum strength inside a nucleus.

Proceeding with this proper choice, and through the use of Eq. (7), one gets

$$E_F = 0 \;, \quad V_F(r) = -\frac{(4\lambda a^2) e^{2a(r-R)}}{\left[e^{2a(r-R)} + 1\right]^2} \;, \tag{12}$$

$$E_f = -a^2 \alpha^2 \;, \quad V_f(r) = -\frac{2a^2[(\alpha+\beta)(\alpha-1)]}{\left[e^{2a(r-R)} + 1\right]} + \frac{a^2[(\alpha+\beta)(\alpha+\beta-2)]}{\left[e^{2a(r-R)} + 1\right]^2} \;, \tag{13}$$

for which

$f = (a)^{-1/2}(-1)^{(1-\alpha)/2}(1-s)^{-\alpha/2}(1+s)^{-\beta/2}$ reproduced by Eq. (8) is employed by Eq. (7). Remembering that $e^{2a(r-R)}/(1+e^{2a(r-R)})^2 = -1/(1+e^{2a(r-R)})^2 + 1/(1+e^{2a(r-R)})$, the total potential influencing a single neutron in a nucleus can be given in a more familiar form,

$$V(r) = V_f(r) + V_F(r) = -\frac{V_0}{[e^{2a(r-R)}+1]} - \frac{W e^{2a(r-R)}}{[e^{2a(r-R)}+1]^2} \quad , \quad E = E_F + E_f = -a^2\alpha^2 \quad . \tag{14}$$

At this point, we note that the similar shell model interaction potentials to investigate nuclear structure of heavy nuclei are recently investigated, see e.g. Ref. [14] and the related references therein. With the consideration these works, it is realized that the additional potential arised naturally in our calculations seems a refinement to the standard Woods-Saxon potential since it increases freedom in the surface structure of the potential as in [14] due to the dip in this region produced by a reasonable choice of $W$. It is also well known that the use of similar potentials is quite useful in the calculation of single-particle energy levels because the pure Woods-Saxon potential does not reproduce such levels with enough accuracy.

Obviosly, the strengths of the potential in Eq. (14)

$$V_0 = a^2(\alpha^2 - \beta^2) \succ 0 \quad , \quad W = a^2[(\alpha+\beta)(\alpha+\beta+4n+2)+4n(n+1)] \succ 0 \tag{15}$$

which coincide exactly with the corresponding definitions in [5]. However, for a physically acceptable interaction, the $n-$dependence of the potential term should be shifted to the related energy value. This can be carried out by constructing a relation between the parameters and strengths of the interaction potential. This relation subsequently determines the $n-$dependence of the spectrum which justifies in principle the discussion of [6] regarding the connection between the number of bound states and the strength of the additional potential ($W$). For the clarification of this point, we proceed with the definition of $W$ in (15) that, after some elemantary calculus, one sees that

$\alpha + \beta = \sqrt{1+W/a^2} - (2n+1)$, and the substitution of which in the $V_0$ expression above makes clear that

$$\alpha = \frac{1}{2}\left[\sqrt{1+W/a^2} - (2n+1) + \frac{V_0/a^2}{\sqrt{1+W/a^2} - (2n+1)}\right], \tag{16}$$

$$\beta = \frac{1}{2}\left[\sqrt{1+W/a^2} - (2n+1) - \frac{V_0/a^2}{\sqrt{1+W/a^2} - (2n+1)}\right]. \tag{17}$$

Reminding $\alpha, \beta \succ -1$ for the Jacobi polynomials, from the above equations we note that

$\alpha - \beta \succ 0$ leading to a restriction such that $\alpha + \beta \succ 0$ since, from Eq. (15), $V_0 \succ 0$. This physically interesting point imposes that $\sqrt{1+W/a^2} \succ 2n+1$, which determines certainly the number of physically meaningful bound states for a deep potential appearing near the surface. Otherwise, $\alpha + \beta \leq 0$ taking to $V_0 \leq 0$ that is not acceptable for the bound state consideration. The second constraint arises due to Eq. (17) where the difference between the two terms in the bracket should be greater than $-2$ in order to validate $\beta \succ -1$. This causes, however, exact calculation of single-particle energy levels lying only in the well generated by $W \succ V_0$ around the surface of the nucleus, since $E_n \succ -V_0$ in this case. This simple analysis, nevertheless, stresses that $W$ cannot disappear in (14) though it seems straightforward in the calculation scheme.

Furthermore, for practical calculations, one needs to visualize a relation between the values of the two potential strengths, since $V_0$ can be defined easily from the literature, see e.g. [15], depending on the total nucleon number of the nucleus considered. To suggest a way of defining proper $W$ values used in the calculations, we remind to the reader that the minimum point of the total potential function in Eq. (14) leads directly to $W$ values. Hence, from $[\partial V(r)/\partial r]_{r=r_{min}} = 0$, $W = \kappa V_0$ where $\kappa = (1+e^{2a(r_{min}-R)})/(1-e^{2a(r_{min}-R)})$. As $W \succ 0$, we have a constraint that $1 \succ e^{2a(r_{min}-R)}$ yielding $r_{min} \prec R(= r_0 A^{1/3})$. The choice of $r_{min}$ location in fact affects the number of bound single-particle states in the nucleus since $r_{min} = R + \ln[(W-V_0)/(W+V_0)]/2a$. For instance, when $r_{min}$ approaches to $R$ the depth of the full potential in (14) increases reasoning an increase in the bound state number, of course, satisfying $\sqrt{1+W/a^2} \succ 2n+1$ restriction.

Bearing in mind these two significant discussions, the exact energy spectrum of the extended Woods-Saxon potential in (14) is

$$E_n = -a^2\alpha^2 = -\frac{a^2}{4}\left\{\left[\sqrt{1+W/a^2}-(2n+1)\right]^2 + \left[\frac{V_0/a^2}{\sqrt{1+W/a^2}-(2n+1)}\right]^2\right\} - \frac{V_0}{2}, \qquad (18)$$

which are consistent with the works in [5,6]. Additionaly, from Eq. (2), the corresponding unnormalized wavefunctions are in the form

$$\Psi_n(r) \propto \left[\frac{1}{e^{2a(r-R)}+1}\right]^{\alpha/2} \left[1-\frac{1}{e^{2a(r-R)}+1}\right]^{\beta/2} P_n^{(\alpha,\beta)} \quad . \tag{19}$$

The expression in Eq. [19] agrees completely with the algebraic definition in [4] where the wavefunction of interest is given in terms of the hypergeometric function $_2F_1(\varepsilon, \varepsilon+1, \alpha+1;(1-s))$, where $\varepsilon = (\alpha+\beta)/2$, instead of $P_n^{(\alpha,\beta)}$ describing excited quantum states $(n \succ 0)$.

Moreover, to test the reliability of Eqs. (15) and (18) we go back to Eq. (14) and remind that this refined form of the Woods-Saxon potential in the $r \succ R$ domain reduces to the usual Morse potential if $a$ is large, which means from the physics point of view that the real diffuseness parameter $a_0$ goes to zero. Whereas in the other region where $r \prec R$, with the same parameter condition, it represents a simple quantum well potential. These testing cases would clarify certainly the credibility of the present calculation results.

Starting with the first case $(r \succ R)$, the potential in (14) is transformed to the form

$$V(r) \to -(V_0+W)e^{-2a(r-R)} + W e^{-4a(r-R)} \quad , \tag{20}$$

which is the Morse like potential. The algebraic form of the solutions for this potential are [3]

$$E_n^{Morse} = -\left[\left(\frac{V_0+W}{2\sqrt{W}}-a\right)-2na\right]^2 \quad , \quad \Psi_n^{Morse} \propto e^{\left[-\left(\frac{V_0+W}{2W}-a\right)(r-R)+\frac{\sqrt{W}}{2a}e^{-2a(r-R)}\right]} F\left(y=e^{-2a(r-R)}\right) \tag{21}$$

where $F$ is the Kummer's standard form of the confluent hypergeometric function related to the Jacobi polynomials [1,2]. The comparison of the wave functions in Eqs. (21) and (19) for the large $a$ value yields

$$\left(\frac{V_0+W}{2\sqrt{W}}-a\right) = \alpha\, a = na \, . \tag{22}$$

At this stage we should point out that, through Eq. (16), obviously $\alpha \to -n$ in case $a \to \infty$. However, for a realistic consideration, $\alpha \succ -1$ due to the properties of $P_n^{(\alpha,\beta)}$ [1], which is in fact satisfied certainly by (16) in its present form except this extreme case ($a \to \infty$) of interest undertaken for testing purposes. Therefore, in order to have physically meaningful results the absolute value of $\alpha$ should be used in such considerations. Hence, the substitution of (22) in (21) leads to a simple energy expression corresponding to the potential in (20), $E_n^{Morse} = -(na)^2$. For the completeness, we re-consider Eq. (18) that reduces to the same energy expression $E_n = -a^2\alpha^2 \rightarrow -a^2 n^2$ for the large $a$ values.

As the second consideration, we investigate the same problem in the $R \succ r$ interaction region where the extended Woods-Saxon potential behaves like a finite quantum well $V(r) = -V_0$ which is one of the well known examples of the quantum theory. The related bound-state wavefunctions inside the nucleus $(R \succ r)$ having a diameter of $2R$ and the corresponding approximated energy values for the low lying states are given as

$$\Psi_n^{FQW} \propto e^{\pm ik\bar{r}} \quad , \quad k = \sqrt{V_0 + E_n^{FQW}} \quad , \quad E_n^{FQW} \approx \frac{n^2 \pi^2}{(2R)^2} - V_0 \quad , \quad E_n^{FQW} \prec 0 \quad . \tag{23}$$

From Eq. (19), the expression for the quantum state wavefunctions of the extended Woods-Saxon potential in the case of large $a$ values reduces to

$$\Psi_n \rightarrow e^{-a\beta\bar{r}} \quad , \quad \bar{r} = R - r \succ 0 \quad , \tag{24}$$

the comparison of which with that of the usual finite well in (23) reveals that $ik = a\beta$. Indeed, this is in well agrement with our one of the earlier definitions, $V_0 = a^2(\alpha^2 - \beta^2)$ leading to, from Eq. (18),

$$k^2 = V_0 - a^2\alpha^2 = V_0 + E_n \quad , \tag{25}$$

which overlaps with the definition of $k$ in (23), justifying the inter-connection between the finite square quantum well and the Woods-Saxon potential in this special case.

This brief comparative study confirms the reliability of present results in a concrete manner. Nevertheless, it is remarked that we have not solved in fact the pure Woods-Saxon potential which is the case if $W = 0$ in Eq. (14). Unfortunately, from the short analysis done above, there is a physical restriction that $W \neq 0$ which has a significant meaning such that the refined potential in (14) cannot be reduced to its well known naked form. In other word, perhaps it means that this potential cannot be solved analytically within the non-relativistic frame like the usual exponential potential. There are some similar examples in the literature representing their exact solutions in the relativistic domain although they have no solution in an explicit form within the non-relativistic framework, such as the exponential potential which is transformed easily to the exactly solvable Morse like potential with the relativistic contributions [16]. Within this context, we make a brief search in the following section to clarify whether the second piece with $W$ of the so called extended or generalized Woods-Saxon potential in (14) is in fact the relativistic correction or not, since it is analytically solvable in its present form.

## 3.2. Relativistic consideration

This section involves an attempt, in the light of [7], to extend the same scenario to the relativistic region to be able to clarify physically interesting point arised naturally during the calculation process mentioned above. For simplicity, only the $s$–wave Klein-Gordon equation leading to bound states is considered.

Among the advantages of working with the potential of interest one-dimensional Klein-Gordon and Dirac equations are solvable in terms of special functions and therefore the study of bound states and scattering processes are more tractable. Because of this, from different perspectives, recently the relativistic Woods-Saxon potential has been extensively discussed [17] in the literature. However, to the best of our knowledge, such a question undertaken in this section has not been delt with so far. The flexibility of the model used here enables us to separate the relativistic contributions explicitly from the non-relativistic solutions, unlike the available theories, which provides a framework to search such problems under consideration.

In the presence of vector and scalar potentials the (1+1)-dimensional time-independent Klein-Gordon equation for a spinless particle of rest mass $m$ reads $(\hbar = c = 1)$

$$-\psi'' + (m + V_s)^2 \psi = (\varepsilon - V_v)^2 \psi \quad , \tag{26}$$

in which $\varepsilon$ is the relativistic energy of the particle, together with $V_v(r)$ and $V_s(r)$ being the vector and scalar potentials respectively. The full relativistic wave function in (26) can be expressed, as in the case of the Schrödinger equation above, by $\psi(r) = F(s)f(r)$ where $F$ now for this specific example denotes the behaviour of the wave function in the non-relativistic region while $f$ represents the modification function due to the relativistic effects. This transformation reproduces similar results to those in the previous section with some slight differences such as the one that $E$ and $V$ terms in Eqs. (1-6) now turn out to be

$$E \to \varepsilon^2 - m^2 \quad , \quad V \to 2(mV_s + \varepsilon V_v) + (V_s^2 - V_v^2) \quad , \tag{27}$$

leading to

$$-\frac{\lambda}{\sigma} s'^2 = 2(mV_s + \varepsilon V_v) - \varepsilon_F \quad , \quad \frac{f''}{f} = (V_s^2 - V_v^2) - \varepsilon_f \quad , \quad \varepsilon^2 - m^2 = \varepsilon_F + \varepsilon_f \quad . \tag{28}$$

Here, $\varepsilon_F$ and $\varepsilon_f$ represent, respectiveley, the energy in the non-relativistic limit and correction to the energy due to the relativistic consideration. For the calculation of $f$ in (28), one should use Eq. (8). The first part in Eq. (28) is the exact appearence of the K-G equation in the non-relativistic domain [7], which can be expressed in terms of orthogonal polynomials, whereas the second equality in there generates the relativistic modifications via a properly constructed $f-$function. If the scaler and vector potentials are equal to each other $(V_s = \pm V_v)$, the relativistic corrections die away and $\psi \to F$ since $f \to$ constant.

Let us proceed with the plausible choice of the scaler and vector potential shapes representing a realistic mean field central potential in nuclei,

$$V_s = -\frac{A}{e^{2a(r-R)}+1} \quad , \quad V_v = -\frac{B}{e^{2a(r-R)}+1} \quad , \tag{29}$$

in which the coupling constants $A$ and $B$ are dimensionless real parameters. Substituting Eq. (29) into Eq. (28) and keeping the previous calculations in mind, Eqs. (9-13), we have

$$V_F(r) = \frac{-2(mA+\varepsilon B)}{e^{2a(r-R)}+1} \quad , \quad V_f(r) = \frac{A^2-B^2}{\left[e^{2a(r-R)}+1\right]^2} = -\frac{(A^2-B^2)e^{2a(r-R)}}{\left[e^{2a(r-R)}+1\right]^2} + \frac{A^2-B^2}{e^{2a(r-R)}+1} \quad , \tag{30}$$

Thus, the total potential $V(r) = V_F + V_f$ and the corresponding state function $\psi = Ff$ that is identical to Eq. (19), together with the relativistic energy spectrum $\varepsilon^2 - m^2 = \varepsilon_f (= -a^2\alpha^2)$, can be readily calculated where

$$\alpha = \frac{1}{2}\left\{-(2n+1) + \sqrt{1+(A^2-B^2)/a^2} + \frac{[2(mA+\varepsilon B)-(A^2-B^2)]/a^2}{\sqrt{1+(A^2-B^2)/a^2} - (2n+1)}\right\} \quad , \tag{31}$$

which agree with those in [17]. Moreover, a careful comparison of $V = V_F + V_f$ with its non-relativistic counterpart in Eq. (14) reveals that $W = A^2 - B^2$ and $V_0 = 2(mA+\varepsilon B) - W$, where $\varepsilon \prec 0$, $A \succ B$ and $2(mA+\varepsilon B) \succ W$.

From this brief search we conclude that the additional potential with the strength of $W$ appearing naturally in the non-relativistic calculations, Eq. (14), is in fact the relativistic contributions as in [16], which disappear if $A = B$ (note that from (29) the choice of $A = -B$ seems physically impossible). However, in this case the potential $V_F(r)$ within the non-relativistic limit reduces to zero as well since $\varepsilon \to m$ in this domain. This result supports why *W* cannot be absent in the non-relativistic treatment, justifying the related discussion presented in the previous section.

## 4. Concluding Remarks

The $s-$wave radial Schrödinger equation with a refined Woods-Saxon potential increasing freedom in the surface structure has been solved exactly in terms of the Jacobi polynomials for the calculation of single neutron energy levels and corresponding wave functions in a nucleus, including deeply bound states. The analytical results obtained have been analysed carefully and the prescriptions used in the present formalism have been discussed in detail, which reveal the reliability and success of the novel treatment for such calculations.

To understand the physics behind the additional potential leading to the exact solvability of the whole potential, the procedure used has been extended successfully for the Klein-Gordon equation. This application has provided a framework to visualise explicitly the solutions at the non-relativistic limit and the corrections due to relativistic consideration. From the consideration of the same problem within the frameworks of the relativistic and non-relativistic physics, we have concluded that the usual Woods-Saxon potential with zero angular momentum cannot be solved analytically. However, its relativistic treatment leads to exact solvability due to the additional potential resembling relativistic like modifications. Clearly considerable additional work is still needed to test further the virtues of the frequently used Woods-Saxon potential description of the nuclear interactions. The work along this line is in progress.